\documentclass[conference,10pt]{IEEEtran}

\usepackage{algorithm}
\usepackage{algorithmic}

\usepackage{cite}
\usepackage{epsfig}
\usepackage{epstopdf}
\usepackage{graphicx}
\usepackage{xcolor}
\usepackage{tikz}
\usetikzlibrary{patterns}
\usetikzlibrary{arrows,positioning,shapes.geometric,circuits.logic.US,spy}
\usepackage{pgfplots}
\usepackage{multirow}
\usepackage{upgreek}
\usepackage{amssymb}
\usepackage{amsmath}
\usepackage{cases}
\usepackage{url}
\usepackage{pifont}
\usepackage{bm}
\usepackage{amsthm}

\newtheorem{proposition}{Proposition}
\newtheorem{lemma}{Lemma}

\usepackage{tabularx}
\usepackage{booktabs}
\usepackage{filecontents}
\usepackage{subcaption}
\newcolumntype{Y}{>{\centering\arraybackslash}X}

\usepackage{array}

\newcommand{\fixme}[2]{\ifx&#2&{\leavevmode\color{red}#1}\else{\leavevmode\color{red}FIXME\{}#1{\leavevmode\color{red}\}}\footnote{{\leavevmode\color{red}#2}}\PackageWarning{Fixme}{#1: #2}\fi}

\newcommand{\newstuff}[2]{\ifx&#2&{\leavevmode\color{blue}#1}\else{\leavevmode\color{blue}FIXME\{}#1{\leavevmode\color{blue}\}}\footnote{{\leavevmode\color{blue}#2}}\PackageWarning{Newstuff}{#1: #2}\fi}

\DeclareMathOperator*{\row}{row}
\DeclareMathOperator*{\vect}{vec}

\title{Construction and Decoding of Product \\ Codes with Non-Systematic Polar Codes}
\author{\IEEEauthorblockN{ Valerio~Bioglio, Carlo~Condo, Ingmar~Land\\}
\IEEEauthorblockA{Mathematical and Algorithmic Sciences Lab\\ Huawei Technologies France SASU \\
Email: $\{$valerio.bioglio,carlo.condo,ingmar.land$\}$@huawei.com}} 
\begin{document}

\maketitle
\begin{abstract}
Product codes are widespread in optical communications, thanks to their high throughput and good error-correction performance.
Systematic polar codes have been recently considered as component codes for product codes. 
In this paper, we present a novel construction for product polar codes based on non-systematic polar codes. 
%We prove that the resulting code is both a product code and a polar code, having a frozen set that is dependent on the frozen sets of the component polar codes. 
We prove that the resulting product code is actually a polar code, having a frozen set that is dependent on the frozen sets of the component polar codes. 
We propose a low-complexity decoding algorithm exploiting the dual nature of the constructed code. 
Performance analysis and simulations show high decoding speed, that allows to construct long codes while maintaining low decoding latency. 
The resulting high throughput and good error-correction performance are appealing for optical communication systems and other systems where high throughput and low latency are required. 

%We propose a low-complexity decoder for the proposed product polar code able to substantially reduce the decoding latency for very long codes, making the proposed construction appealing for optical communication systems.
\end{abstract}

%\begin{IEEEkeywords}
%
%\end{IEEEkeywords}

\IEEEpeerreviewmaketitle

\section{Introduction} \label{sec:intro}

Polar codes \cite{arikan} are capacity-achieving linear block codes based on the polarization phenomenon, that makes bit channels either completely noisy or completely noiseless as code length tends to infinity. 
While optimal at infinite code length, the error-correction performance of polar codes under successive cancellation (SC) decoding degrades at practical code lengths. 
Moreover, SC-based decoding algorithms are inherently sequential, which results in high dependency of decoding latency on code length.
List decoding was proposed in \cite{tal_list} to improve SC performance for practical code lengths: the resulting SC-List (SCL) algorithm exhibits enhanced error-correction performance, at the cost of higher decoder latency and complexity. 
%SC-based decoding algorithms are inherently sequential, and can result in long decoding latency.
%enhanced performance of this algorithm contributed to the selection of polar codes as a coding scheme for 5G wireless standards.

Product codes \cite{elias} are parallel concatenated codes often used in optical communication systems for their good error-correction performance and high throughput, thanks to their highly parallelizable decoding process.
To exploit this feature, systematic polar codes have been concatenated with short block codes as well as LDPC codes \cite{pc_inner,BP_pc}.
This concatenation allows the construction of very long product codes based on the polarization effect: to fully exploit the decoding parallelism, a high number of parallel decoders for the component codes need to be instantiated, leading to a high hardware cost.
%Decoding of product codes uses iterative hard- or soft-decision decoding \cite{Jian_PROD,Sugihara_PROD}.
%\fixme{Short block codes as well as LDPC codes have been concatenated with polar codes in order to improve the throughput of long polar codes while keeping good decoding performance \cite{pc_inner,BP_pc}.}{Not sure how this improves the throughput.}
%However, these concatenation schemes exhibit poor error correction performance if compared to LDPC or turbo codes. 
%Authors in \cite{par_conc_sys} propose to use systematic polar codes in concatenation scheme with recursive systematic convolutional codes to reduce burst errors probability. 
Authors in \cite{par_conc_sys_pol} propose to use two systematic polar codes in the concatenation scheme in order to simplify the decoder structure. 
Soft cancellation (SCAN) \cite{SCAN_pc} and belief propagation (BP) \cite{BP_pc} can be used as soft-input / soft-output decoders for systematic polar codes, at the cost of increased decoding complexity  compared to SC.
Recently, SCL decoding has been proposed as a valid alternative to SCAN and BP \cite{par_sys_list}, while authors in \cite{KoikeAkinoIrregularPT} propose to use irregular systematic polar codes to further increase the decoding throughput. 
%\fixme{Parallel concatenated systematic polar codes have been tested as error-correction codes for magnetic recording technologies, showing good performance compared to state-of-the-art turbo-equalized architectures based on LDPC codes \cite{pc_magnetic}. }{Probably unnecessary}

In this paper, we show that the nature of polar codes inherently induces the construction of product codes that are not systematic. 
%We show that the concatenation of two polar codes is a polar code that can be designed and decoded as a product code. 
In particular, we show that the product of two polar codes is a polar code, that can be designed and decoded as a product code. 
%This property allows us to construct product polar codes more efficiently. 
We propose a code construction approach and a low-complexity decoding algorithm that makes use of the observed dual interpretation of polar codes. 
Both analysis and simulations show that the proposed code construction and decoding approaches allow to combine high decoding speed and long codes, resulting in high-throughput and good error-correction performance suitable for optical communications. % systems. 

\section{Preliminaries} \label{sec:prel}

\subsection{Polar Codes} \label{subsec:PC}

Polar codes are linear block codes based on the polarization effect of the kernel matrix $T_2 = \left[\begin{smallmatrix} 1&0\\1&1 \end{smallmatrix}\right]$.
A polar code of length $N=2^n$ and dimension $K$ is defined by the transformation matrix $T_N = T_2 ^{\otimes n}$, given by the $n$-fold Kronecker power of the polarization kernel, and a frozen set $\mathcal{F} \subset \{1,\dots,N\}$ composed of $N-K$ elements.
Codeword $x = [x_0,x_1,\ldots,x_{N-1}]$ is calculated as 
%Polar codes are linear block codes with code length $N=2^n$ and code rate $R = K/N$. They are constructed through the transformation matrix $D^{\otimes n}$ as
\begin{equation}
x = u \cdot T_N \text{,} \label{eq:polarGen}
\end{equation}
%$x = u \cdot T_N \text{,}$
%where $x = \{x_0,x_1,\ldots,x_{N-1}\}$ is the coded vector and 
where the input vector $u = [u_0,u_1,\ldots,u_{N-1}]$ has the $N-K$ bits in the positions listed in $\mathcal{F}$ set to zero, while the remaining $K$ bits carry the information to be transmitted. 
%is the input vector, constituted of $K$ information bits and $N-K$ frozen bits listed in $\mathcal{F}$, whose value is usually set to zero and is known by the decoder. 
%Matrix $D^{\otimes n}$ is the $n$-th Kronecker product of the polarization kernel $D = \left[\begin{smallmatrix} 1&0\\1&1 \end{smallmatrix}\right]$.
%The polarization process allows to sort the $N$ bits positions of $u$ in order of reliability. Consequently, the $K$ information bits are assigned to the most reliable bit-channels. The set of bit channels identifying the frozen bits constitutes the frozen set $\mathcal{F}$. 
%As $N$ tends to infinity, the polarization phenomenon makes bit channels either completely noisy or completely noiseless, where the fraction of noiseless bit-channels equals the channel capacity \cite{arikan}. 
The frozen set is usually designed to minimize the error probability under SC decoding, such that information bits are stored in the most reliable bits, defining the information set $\mathcal{I} = \mathcal{F}^C$. 
Reliabilities can be calculated in various ways, e.g. via Monte Carlo simulation, by tracking the Batthacharyya parameter, or by density evolution under a Gaussian approximation \cite{polar_const}. 
The generator matrix $G$ of a polar code is calculated from the transformation matrix $T_N$ by deleting the rows of the indices listed in the frozen set. 

%\fixme{Ho messo il paragrafo seguente, ma complica un po' la notazione.}{}
%Given a binary memoryless symmetric channel $W$ be a binary memoryless symmetric channel, the reliability of the channel can be expressed through the Bhattacharyya parameter $Z(W)\in [0,1]$, that is defined as
%\begin{align}\label{eq:Battapar}
%& Z(W)= \sum_{y \in \mathcal{Y}} \sqrt{W(y\mid 0)W(y \mid 1)}~,
%\end{align}
%where $\mathcal{X}$ and $\mathcal{Y}$ are the input and output alphabets, and $\{W(y \mid x) : x\in \mathcal{X}, y\in \mathcal{Y}\}$ the transition probabilities. Reliable bit-channels are characterized by a low Bhattacharyya parameter.
%For finite code lengths, however, the polarization effect is incomplete. 
%The polar encoding process consists in the classification of the bit-channels in $u$ into two groups: the $K$ reliable bit-channels, to which information set $\mathcal{I}$ is assigned, and the $N-K$ unreliable bit-channels, that are set to $0$ and are indexed by the frozen set $\mathcal{F}$. 

SC decoding \cite{arikan} can be interpreted as a depth-first binary tree search with priority given to the left branches. 
Each node of the tree receives from its parent a soft information vector, that gets processed and transmitted to the left and right child nodes. 
Bits are estimated at leaf nodes, and hard estimates are propagated from child to parent nodes. 
While optimal for infinite codes, SC decoding exhibits mediocre performance for short codes. 
%suffers from 
SCL decoding \cite{tal_list} maintains $L$ parallel codeword candidates, improving decoding performance of polar codes for moderate code lengths. 
%: when an information bit is estimated, the number of paths doubles, and $L$ candidates are discarded thanks to a path metric, improving decoding performance of polar codes for moderate code lengths. 
The error-correction performance of SCL can be further improved by concatenating the polar code with a cyclic redundancy check (CRC), that helps in the selection of the final candidate. 
%Improvements for SC-based decoding algorithms have been proposed in \cite{sarkis,hashemi_SSCL,hashemi_FSSCL}. 
%They rely on the identification of particular frozen bit patterns to prune the decoding tree and reduce the decoding latency.

\subsection{Product Codes} \label{subsec:prod}
Product codes were introduced in \cite{elias} as a simple and efficient way to build very long codes on the basis of two or more short block component codes. 
Even if it is not necessary, component codes are usually systematic in order to simplify the encoding. 
In general, given two systematic linear block codes $\mathcal{C}_r$ and $\mathcal{C}_c$ with parameters $(N_r,K_r)$ and $(N_c,K_c)$ respectively, the product code $\mathcal{P} = \mathcal{C}_c \times \mathcal{C}_r$ of length $N = N_r N_c$ and dimension $K = K_r K_c$ is obtained as follows. 
The $K$ information bits are arranged in a $K_c \times K_r$ matrix $U$, then code $\mathcal{C}_r$ is used to encode the $K_c$ rows independently. 
Afterwards, the $N_r$ columns obtained in the previous step are encoded independently using code $\mathcal{C}_c$. 
The result is a $N_c \times N_r$ codeword matrix $X$, where rows are codewords of code $\mathcal{C}_r$ and columns are codewords of code $\mathcal{C}_c$, calculated as 
%This procedure can be written as %is summarized by equation
\begin{equation}
X = G_c^T \cdot U \cdot G_r,
\end{equation}
where $G_r$ and $G_c$ are the generator matrices of codes $\mathcal{C}_r$ and $\mathcal{C}_c$ respectively. 
Alternatively, the generator matrix of $\mathcal{P}$ can be obtained taking the Kronecker product of the generator matrices of the two component codes as $G = G_c \otimes G_r$ \cite{mcw_sloane}. 
%Let $G_r$ and $G_c$ be the generator matrices of the codes $\mathcal{C}_r$ and $\mathcal{C}_c$ respectively, then the generator matrix $G$ of the product code $\mathcal{P}$ is equal to $G = G_c \otimes G_r$.

Product codes can be decoded by sequentially decoding rows and column component codes, and exchanging information between the two phases. % in order to reduce the decoding complexity. 
Soft-input/soft-output algorithms are used to improve the decoding performance by iterating the decoding of rows and columns and exchanging soft information between the two decoders \cite{block_turbo}. 
Since no information is directly exchanged among rows (columns), the decoding of all row (column) component codes can be performed concurrently.

%\section{Polar Codes as Product Codes} \label{sec:prodPol}
\section{Product Polar Codes Design} \label{sec:prodPol}
\begin{figure}
  \centering
  \includegraphics[width=0.45\textwidth]{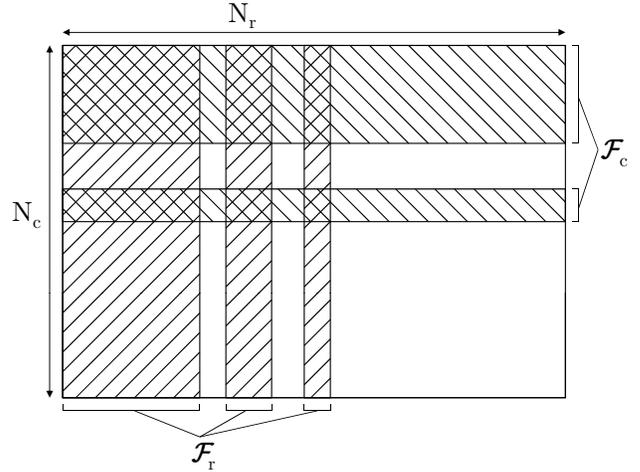}
  \caption{Input matrix $U$ for a product polar code.}
  \label{fig:prod_pol_des}
\end{figure}
Product codes based on polar codes have been proposed in literature, using systematic polar codes as one of the two component codes or as both. 
However, the peculiar structure of polar codes has never been exploited in the construction of the product code. 
%How the frozen set of the length-$N$ code is bound to that of the length-$\sqrt{N}$ codes.
Both polar and product codes are defined through the Kronecker product of short and simple blocks, that are used to construct longer and more powerful codes. 
In the following, we prove that the product of two non-systematic polar codes is still a polar code, having a peculiar frozen set obtained on the basis of the component polar codes. 
%Definition of product codes is really similar to the definition of polar codes. 
%In fact, both codes allow to construct long codes on the basis of short and simple blocks, and both are based on the Kronecker product of the generator matrices of the constituent codes. 
%In the following, we design and decode polar codes constituted as a product of two shorter polar codes. 
%This motivate us to design polar codes constituted as a product of two shorter polar codes. 
This design can be extended to multi-dimensional product codes. 

%In this paper, we want to show that the nature of polar codes permits to build product codes without the need of systematic polar codes. 
%More precisely, we will show that polar codes are actually product codes, and can be designed and decoded as product codes. 

%\subsection{Definition of the frozen set} \label{subsec:F}
%\subsection{Product Polar Code Design} \label{subsec:F}
%%
%\begin{figure}
%  \centering
%  \includegraphics[width=0.40\textwidth]{figures/prod_scheme.eps}
%  \caption{Product code scheme.}
%  \label{fig:scheme}
%\end{figure}
%%

Let us define two polar codes $\mathcal{C}_r$ and $\mathcal{C}_c$ of parameters $(N_r,K_r)$ and $(N_c,K_c)$ with transformation matrices $T_{N_r}$ and $T_{N_c}$ respectively, where $N_c = 2^{n_c}$ and $N_r = 2^{n_r}$, and $\mathcal{F}_r$ and $\mathcal{F}_c$ are the respective frozen sets. 
%The structure of the proposed scheme is depicted in Figure~\ref{fig:scheme}. 
The product polar code $\mathcal{P} = \mathcal{C}_c \times \mathcal{C}_r$ is generated as follows. 
An $N_c \times N_r$ input matrix $U$ is generated having zeros in the columns listed in $\mathcal{F}_r$ and in the rows listed in $\mathcal{F}_c$ as depicted in Figure~\ref{fig:prod_pol_des}. 
Input bits are stored in the remaining $K_r K_c$ entries of $U$, row first, starting from the top left entry. 
Encoding is performed as for product codes: the rows of $U$ are encoded independently using polar code $\mathcal{C}_r$, namely through matrix multiplication by the transformation matrix $T_{N_r}$, obtaining matrix $U_r$. 
Then, the columns of $U_r$ are encoded independently using $\mathcal{C}_c$. 
The encoding order can be inverted performing column encoding first and row encoding next without changing the results. 
The resulting codeword matrix $X$ can be expressed as
\begin{equation}
X = T_{N_c}^T \cdot U \cdot T_{N_r}.
\end{equation}
%In the following, we show that this procedure creates a polar code. 
In order to show that this procedure creates a polar code, let us vectorize the input and codeword matrices $U$ and $X$, converting them into row vectors $u$ and $x$. 
This operation is performed by the linear transformation $\row(\cdot)$, which converts a matrix into a row vector by juxtaposing its rows head-to-tail. 
This transformation is similar to the classical vectorization function $\vect(\cdot)$ converting a matrix into a column vector by juxtaposing its columns head-to-tail.
However, before proving our claim, we need to extend a classical result of $\vect(\cdot)$ function to $\row$ function.
\begin{lemma}
\label{prop:row}
Given three matrices $A$, $B$, $C$ such that $A \cdot B \cdot C$ is defined, then
\begin{equation}
\row(A \cdot B \cdot C) = \row(B) \cdot (A^T \otimes C).
\end{equation}
\begin{proof}
The compatibility of vectorization with the Kronecker product is well known, and is used to express matrix multiplication $A \cdot B \cdot C$ as a linear transformation $\vect(A \cdot B \cdot C) = (C^T \otimes A) \cdot \vect(B)$. 
Moreover, by construction we have that $\vect(A^T) = (\row(A))^T$. 
As a consequence,
\begin{align*}
\row(A \cdot B \cdot C) & =  (\vect((A \cdot B \cdot C)^T))^T  \\
  & =  (\vect(C^T \cdot B ^T\cdot A^T))^T  \\
  & =  ((A \otimes C^T) \cdot \vect(B^T))^T  \\
  & =  (\vect(B^T))^T \cdot  (A \otimes C^T)^T  \\
  & =  \row(B) \cdot (A^T \otimes C).
\end{align*}
\end{proof}
\end{lemma}
Equipped with Lemma~\ref{prop:row} we can now prove the following proposition: 
\begin{proposition}
\label{prop:frozen}
The $(N,K)$ product code $\mathcal{P}$ defined by the product of two non-systematic polar codes as $\mathcal{P} = \mathcal{C}_c \times \mathcal{C}_r$ is a non-systematic polar code having transformation matrix $T_N = T_{N_c} \otimes T_{N_r}$ and frozen set %$\mathcal{F}$ given by
%polar code produced by the product of two polar codes $\mathcal{C}_c$ and $\mathcal{C}_r$ is equal to 
%with frozen sets $\mathcal{F}_c$ and $\mathcal{F}_r$ is equal to
%\[ \mathcal{F} = \mathcal{F}_c \cup \mathcal{F}_r \]
\begin{equation}
\label{eq:frozen}
\mathcal{F} = \arg \min (i_c \otimes i_r) ,
\end{equation} 
where $i_r$ ($i_c$) is a vector of length $N_r$ ($N_c$) having zeros in the positions listed in $\mathcal{F}_r$ ($\mathcal{F}_c$) and ones elsewhere. 
\begin{proof}
%By construction, product polar code $\mathcal{P}$ has transformation matrix $T_N = T_{N_c} \otimes T_{N_r}$ due to the definition of product codes and the fact that the generator matrix of a polar code is a sub-matrix of its transformation matrix. 
% given by the rows listed in the information set, while the generator matrix of the product code $\mathcal{P}$ is $G = G_c \otimes G_r$. 
To prove the proposition we have to show that $x = \row(X)$ is the codeword of a polar code, providing its frozen set and transformation matrix. 
If $u = \row(U)$, Lemma~\ref{prop:row} shows that 
\begin{align*}
x & = \row(X) \\
  & = \row(T_{N_c}^T \cdot U \cdot T_{N_r}) \\
  & = \row(U) \cdot (T_{N_c} \otimes T_{N_r}) \\
  & = u \cdot T_N. 
\end{align*}
%By construction, input vector $u$ has a frozen set imposed by the frozen set \eqref{eq:frozen} of input matrix $U$. 
By construction, input vector $u$ has zero entries in positions imposed by the structure of the input matrix $U$, and (\ref{eq:frozen}) follows from the definition of $U$; with a little abuse of notation, we use the $\arg \min$ function to return the set of the indices of vector $i = i_c \otimes i_r$ for which the entry is zero. 
Finally, $T_N = T_{N_c} \otimes T_{N_r} = T_2^{\otimes(n_c + n_r)}$ is the transformation matrix of a polar code of length $N = 2^{n_c + n_r}$.
\end{proof}
\end{proposition}

Proposition~\ref{prop:frozen} shows how to design a product polar code on the basis of the two component polar codes. 
%Given two polar codes $\mathcal{C}_c$ and $\mathcal{C}_r$ of parameters $(N_c,K_c)$ and $(N_r,K_r)$, with frozen sets $\mathcal{F}_c$ and $\mathcal{F}_r$, it is possible to create a polar code $\mathcal{P}$ of parameters $(N,K)$ and frozen set $\mathcal{F}$ that is the product of the two polar codes, with $N = N_r N_c$ and $K = K_r K_c$. 
The resulting product polar code $\mathcal{P}$ has parameters $(N,K)$, with $N = N_r N_c$ and $K = K_r K_c$, and frozen set $\mathcal{F}$ designed according to \eqref{eq:frozen}. 
The encoding of $\mathcal{P}$ can be performed in $O(\log N)$ steps exploiting the structure of $T_N$. 
%If the obtained codeword $x$ is rearranged in the $N_c \times N_r$ matrix $X$, the sub-vectors $x_r^i$ and $x_c^j$ corresponding to the $i$-th row and the $j$-th column of $X$ represent codewords of polar codes $\mathcal{C}_r$ and $\mathcal{C}_c$ respectively. 
The sub-vectors $x_r^i$ and $x_c^j$ corresponding to the $i$-th row and the $j$-th column of $X$ represent codewords of polar codes $\mathcal{C}_r$ and $\mathcal{C}_c$ respectively. 
%Obtained codeword $x$ can be rearranged in an $N_c \times N_r$ array as depicted in Figure~\ref{fig:scheme}. 
%If we call $x_r^i$ and $x_c^j$ the $i$-th row and the $j$-th column of this structure, these represent codewords of polar codes $\mathcal{C}_r$ and $\mathcal{C}_c$ respectively. 
It is worth noticing that the frozen set identified for the product polar code is suboptimal, w.r.t. SC decoding, compared to the one calculated for a polar code of length $N$. 
On the other hand, this frozen set allows to construct a polar code as a result of the product of two shorter polar codes, that can be exploited at decoding time to reduce the decoding latency, as shown in Section~\ref{subsec:lat}. 
We also conjecture the possibility to invert the product polar code construction, decomposing a polar code as the product of two or more shorter polar codes. 

\begin{figure}
  \centering
  \includegraphics[width=0.25\textwidth]{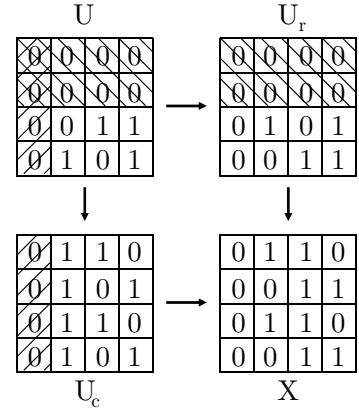}
  \caption{Example of product polar code design and encoding.}
  \label{fig:prod_pol_ex}
\end{figure}
Figure~\ref{fig:prod_pol_ex} shows the encoding of a product polar code generated by a $(4,2)$ polar code with frozen set $\mathcal{F}_c = \{ 0,1 \}$ as column code $\mathcal{C}_c$ and a $(4,3)$ polar code with frozen set $\mathcal{F}_r = \{ 0 \}$ as row code $\mathcal{C}_r$. 
This defines a product polar code $\mathcal{P}$ with $N = 16$ and $K = 6$. 
According to Proposition~\ref{prop:frozen}, its frozen set can be calculated through the Kronecker product of the auxiliary vectors $i_c = [0,0,1,1]$ and $i_r = [0,1,1,1]$, obtaining $\mathcal{F} = \{ 0,1,2,3,4,5,6,7,8,12 \}$. 
We recall that the optimal frozen set for a $(16,6)$ polar code is given by $\mathcal{F}' = \{ 0,1,2,3,4,5,6,8,9,10 \}$. 

%It is worth to notice that it is possible also to invert the construction order, namely taking a polar code and decompose it as the product of smaller polar codes. 
%This process is out of the scope of this paper and will be elaborated in a future paper.

%\subsection{Proposed Two-Step Decoding} \label{subsec:dec}
\section{Low-latency Decoding of Product Polar Codes} \label{sec:dec}
\begin{figure}
  \centering
  \includegraphics[width=0.45\textwidth]{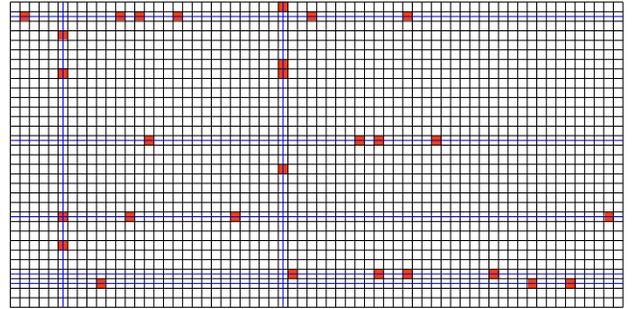}
  \caption{Example of overlapping of $X_r$ and $X_c$. Red squares represent mismatches, blue lines represent wrong estimations identified by Algorithm \ref{alg:find_err}.}
  \label{fig:err_patt}
\end{figure}
%
%Proposed decoding approaches for low-latency/low-complexity product-like decoding.
In this Section, we present a two-step, low-complexity decoding scheme for the proposed polar product codes construction, based on the dual nature of these codes. 
%Code $\mathcal{P}$ can in fact be decoded through SC, however the decoding latency would grow linearly with the code length due the sequential nature of this algorithm. 
We propose to initially decode the code as a product code (step 1), and in case of failure to perform SC decoding on the full polar code (step 2). 
The product code decoding algorithm of step 1 exploits the soft-input / hard-output nature of SC decoding to obtain a low complexity decoder for long codes.  
% has very low complexity compared to SC decoding of the full code, which is 
We then analyze the complexity and expected latency of the presented decoding approach.

\subsection{Two-Step Decoding} \label{subsec:dec}
%In the first step, we propose to decode the component polar codes independently, comparing the results: if the row and column decoders return the same results, decoding stops, otherwise estimations labeled as wrong are attempted again.
The first decoding step considers the polar code as a product code.
Vector $y$ containing the log-likelihood ratios (LLRs) of the $N$ received bits is rearranged in the $N_c \times N_r$ matrix $Y$. 
% depicted in \fixme{Figure~\ref{fig:scheme}}{Necessary?}. 
%We call $x_r^i$ and $x_c^j$ the $i$-th row and the $j$-th column of this structure, that represent codewords of polar codes $\mathcal{C}_r$ and $\mathcal{C}_c$ respectively. 
Every row is considered as a noisy $\mathcal{C}_r$ polar codeword, and decoded independently through SC to estimate vector $\hat{u}_r$.
Each $\hat{u}_r$ is re-encoded, obtaining $\hat{x}_r=\hat{u}_r\cdot T_{N_r}$: the $N_r$-bit vectors $\hat{x}_r$ are then stored as rows of matrix $X_r$. 
The same procedure is applied to the columns of $Y$, obtaining vectors $\hat{x}_c=\hat{u}_c\cdot T_{N_c}$, that are in turn stored as columns of matrix $X_c$. 
In case $X_r = X_c$, decoding is considered successful; the estimated input vector $\hat{u}$ of code $\mathcal{P}$ can thus be derived inverting the encoding operation, i.e. by encoding vector $\hat{x}=\row(X_r)$, since $T_N$ is involutory. %, i.e. via matrix multiplication with the transformation matrix $T_N$. 
%matrix multiplication of the vectorized version of $X_r$ with the transformation matrix $T_N$. 
In case $X_r \neq X_c$, it is possible to identify incorrect estimations by overlapping $X_r$ and $X_c$ and observing the pattern of mismatches.
%Since decoding errors in a row are usually correct in the intersecting columns, and vice-versa, m
Mismatches are usually grouped in strings, as shown in Figure~\ref{fig:err_patt}, where mismatches are represented by red squares.

% then an error occurred in the decoding. 
%This means that either some row or some column have been decoded incorrectly. 
%Overlapping $X_r$ and $X_c$, it is possible to guess these incorrect codewords since differences between the two matrices are grouped in strings, as shown in Figure~\ref{fig:err_patt}. 
%In general, the lower the component codes rate, the higher is the number of mismatches of an incorrect codeword. 

%
\begin{algorithm}[t!]
\caption{FindErroneousEstimations} \label{alg:find_err}
\begin{algorithmic}[1]
%\Procedure{MyProcedure}{}
\STATE Initialize $\text{ErrRows} = \text{ErrCols} = \emptyset$
\STATE $X_d = X_r \oplus X_c$%(X_r \neq X_c)$
\STATE $\text{NumErrRows} = \text{SumRows}(X_d) $
\STATE $\text{NumErrCols} = \text{SumCols}(X_d) $
\WHILE{$\text{NumErrRows} + \text{NumErrCols} > 0$}
	  \STATE $e_r=\text{arg max(NumErrRows)}$ 
  	  \STATE $e_c=\text{arg max(NumErrCols)}$ 
   \IF{$\text{max(NumErrRows)} > \text{max(NumErrCols)}$}
      \STATE $\text{ErrRows} = \text{ErrRows}  \cup \{e_r\}$
      \STATE $X_d(e_r,:) = 0$
   \ELSE
      \STATE $\text{ErrCols} = \text{ErrCols} \cup \{e_c\}$
      \STATE $X_d(:,e_c) = 0$
   \ENDIF
   \STATE $\text{NumErrRows} = \text{SumRows}(X_d) $
   \STATE $\text{NumErrCols} = \text{SumCols}(X_d) $
\ENDWHILE
\RETURN ErrRows, ErrCols
%\EndProcedure
\end{algorithmic}
\end{algorithm}
Even if mismatch patterns are simple to analyze by visual inspection, it may be complex for an algorithm to recognize an erroneous row or column. 
We propose the greedy Algorithm~\ref{alg:find_err} to accomplish this task. 
%detect the incorrect codewords (\textbf{describe algo}) . 
The number of mismatches in each row and column is counted, flagging as incorrect that with the highest count.
Next, its contribution is subtracted from the mismatch count of connected rows or columns, and another incorrect one is identified. 
The process is repeated until all mismatches belong to incorrect rows or columns, the list of which is stored in $\text{ErrRows}$ and $\text{ErrCols}$.
An example of this identification process is represented by the blue lines in Figure~\ref{fig:err_patt}.

Incorrect rows can be rectified using correct columns and vice-versa, but intersections of wrong rows and columns cannot.
In order to correct these errors, we propose to treat the intersection points as erasures.  
As an example, in a row, crossing points with incorrect columns have their LLR set to 0, while intersections with correct columns set the LLR to $+\infty$ if the corresponding bit in $X_c$ has been decoded as $0$, and to $-\infty$ if the bit is $1$.
The rows and columns flagged as incorrect are then re-decoded, obtaining updated $X_r$ and $X_c$.
This procedure is iterated a number $t$ of times, or until $X_r = X_c$.
%In case $X_r \neq X_c$ after $t$ iterations, the second step of the proposed decoding approach is activated. 
%The algorithm used to decode the component codes in step one is used to decode the received vector $y$, considering the complete code $\mathcal{P}$. 

In case $X_r \neq X_c$ after $t$ iterations, the first step returns a failure.
In this case, the second step of the algorithm is performed, namely the received vector $y$ is decoded directly, considering the complete length-$N$polar code $\mathcal{P}$. 

The proposed two-step decoding approach is summarized in Algorithm~\ref{alg:prod_pol}.
Any polar code decoder can be used at lines 3,4 and 16. 
However, since a soft output is not necessary, and the decoding process can be parallelized, simple, sequential and non-iterative SC-based algorithms can be used instead of the more complex BP and SCAN.

\begin{algorithm}[t!]
\caption{TwoStepDecoding} \label{alg:prod_pol}
\begin{algorithmic}[1]
%\Procedure{MyProcedure}{}
\STATE Initialize $Y_r = Y_c = Y $
\FOR{$w = 1 \dots t$}
   \STATE $\hat{X}_r = \text{DecodeRows}(Y)$
   \STATE $\hat{X}_c = \text{DecodeCols}(Y)$
   \IF{$X_r == X_c$}
      \STATE $\hat{x} = \row(X_r)$
      \RETURN $\hat{u} = \text{PolarEncoding}(\hat{x})$
%      \STATE break
   \ELSE
      \STATE FindErroneousEstimations
   \ENDIF 
   \STATE $Y_r = (-2 \hat{X}_c + 1) \cdot \infty$
   \STATE $Y_c = (-2 \hat{X}_r + 1) \cdot \infty$
   \STATE $Y_r(:,\text{ErrCols}) = 0$
   \STATE $Y_c(\text{ErrRows},:) = 0$
\ENDFOR
\RETURN $\hat{u} = \text{Decode}(y)$  
%\EndProcedure
\end{algorithmic}
\end{algorithm}

\subsection{Decoding Latency and Complexity} \label{subsec:lat}

The proposed two-step decoding of product polar codes allows to split the polar decoding process into $N_r+N_c$ shorter, independent decoding processes, whose hard decisions are compared and combined together, using the long polar code decoding only in case of failure.
Let us define as $\Delta_N$ the number of time steps required by an SC-based algorithm to decode a polar code of length $N$. 
For the purpose of latency analysis, we suppose the decoder to have unlimited computational resources, allowing a fully parallel implementation of decoding algorithms. 
Using Algorithm~\ref{alg:prod_pol} to decode component codes, the expected number of steps for the proposed two-step decoder for a code of length $N = N_c N_r$ is given by
%Also, we identify as P-SC the proposed two-step decoding with component codes decoded with SC. P-SC decoding of a code of length $N=N_c\times N_r$ requires 
%\Delta^{\rm P-SC}_N = \left(2\times\max(N_r,N_c)-2\right)\times t_{avg} + \gamma(2N-2)~,
\begin{equation} \label{eq:time}
\Delta^{\rm P}_N = t_{avg} \Delta_{\max(N_r,N_c)} + \gamma \Delta_N~,
\end{equation}
where $t_{avg}\le t$ is the average number iterations, and $\max(N_r,N_c)$ assumes that the decoding of row and column component codes is performed at the same time.
The parameter $\gamma$ is the fraction of decoding attempts in which the second decoding step was performed. 
It can be seen that as long as $\gamma\approx 0$ and $t_{avg}<<N/\max(N_r,N_c)$, then $\Delta^{\rm P}_N$ is substantially smaller than $\Delta_N$.
%P-SC is substantially lower than that of SC decoding. 

The structure of parallel and partially-parallel SC-based decoders is based on a number of processing elements performing LLR and hard decision updates, and on dedicated memory structures to store final and intermediate values. 
Given the recursive structure of polar codes, decoders for shorter codes are naturally nested within decoders for longer codes. 
In the same way, the main difference between long and short code decoders is the amount of memory used. 
Thus, not only a high degree of resource sharing can be expected between the first and second decoding step; the parallelization available during the first decoding step implies that the same hardware can be used in the second step, with minor overhead.

%\textbf{add discussion on hardware implementation: prove that we can use the same resources for both steps}

\section{Performance Results} \label{sec:res}

%\fixme{SHORTEN}{}
%

The dual nature of product polar codes can bring substantial speedup in the decoding; on the other hand, given a time constraint, longer codes can be decoded, leading to improved error-correction performance. 
In this Section, we present decoding speed and error-correction performance analysis, along with simulation results. 
We assume an additive white Gaussian noise (AWGN) channel with binary phase-shift keying (BPSK) modulation, while the two component codes have the same parameters, i.e. $N_r = N_c$ and $K_r = K_c$. 

\begin{figure}
  \centering
  \includegraphics[width=0.47\textwidth]{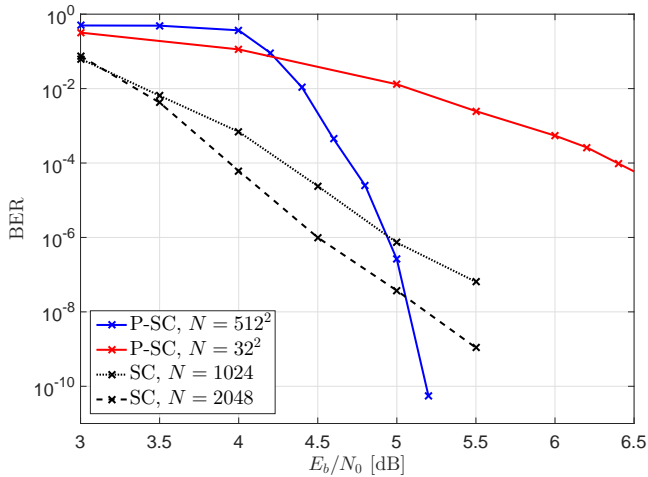}
  \caption{BER comparison for SC and P-SC, for codes of rate $R=(7/8)^2$.}
  \label{fig:ECP-SC}
\end{figure}

\subsection{Error-Correction Performance} \label{subsec:ECP}

As explained in Section~\ref{sec:prodPol}, the frozen set identified for the code of length $N$ is suboptimal for product decoding of polar codes, that relies on the frozen set seen by component codes. 
On the other hand, a frozen set that can help product decoding leads to error-correction performance degradation when standard polar code decoding is applied.

%Fig. \ref{fig:ECP-SC} portrays the bit error rate (BER) for four different codes. 
%The black and magenta curves have been obtained through P-SC decoding of $N=512\times 512$ and $N=32\times32$ polar code, respectively. 
%The component codes have a rate of $R_r=R_c=7/8$. 
%The P-SC decoding approach with $t=4$ has been used, and the frozen set has been selected as the optimal one for $N_r=N_c=512$ and $N_r=N_c=32$, respectively. 
%The red and blue curves have been obtained with SC decoding of a polar code of length $N=1024$, $R=(7/8)^2$ and $N=2048$, $R=(7/8)^2$, the frozen set obtained as \fixme{in \cite{arikan}}{}.
Figure~\ref{fig:ECP-SC} portrays the bit error rate (BER) for different codes under P-SC decoding, i.e. the proposed two-step decoding with SC as the component decoder, with parameter $t=4$, while $N = 512^2 = 262144$ and $N = 32^2 = 1024$ with rate $R = (7/8)^2$. 
As a reference, Figure~\ref{fig:ECP-SC} displays also curves obtained with SC decoding of a polar code of length $N=1024$ and $N=2048$, with the same rate $R=(7/8)^2$, designed according to \cite{arikan}.
As expected due to the suboptimality of the frozen set, P-SC degrades the error correction performance with respect to standard SC decoding when compared to codes with the same code length $N$. 
%At the same time, as detailed in Section \ref{subsec:speed}, P-SC can achieve a substantial speedup over standard SC. 
%Thus, comparing codes with similar decoding speed implies that the code decoded with our proposed approach can have a larger $N$.
However, the speedup achieved by P-SC over standard SC allows to decode longer codes within the same time constraint: consequently, we compare codes with similar decoding latency. %it makes more sense to compare codes with similar decoding latency. 
SC decoding of $N=2048$ and $N=1024$ codes has a decoding latency similar to that of a conservative estimate for P-SC decoding of the $N=512^2$ code. 
The steeper slope imposed by the longer code can thus be exploited within the same time frame as the shorter codes: the BER curves are shown to cross at around $\text{BER} \simeq 10^{-7}$. % and $\text{BER} \simeq 10^{-11}$, depending on the code.

%\begin{figure}
%  \centering
%   \scalebox{1}{\input{figures/ECP1.tikz}}
%   \ref{ECP-SC}
%  \\
%  \vspace{2pt}
%  \caption{BER and BLER curves for SC and P-SC, for codes of rate $R=(7/8)^2$. }
%  \label{fig:ECP-SC}
%\end{figure}

\begin{figure}
  \centering
  \includegraphics[width=0.47\textwidth]{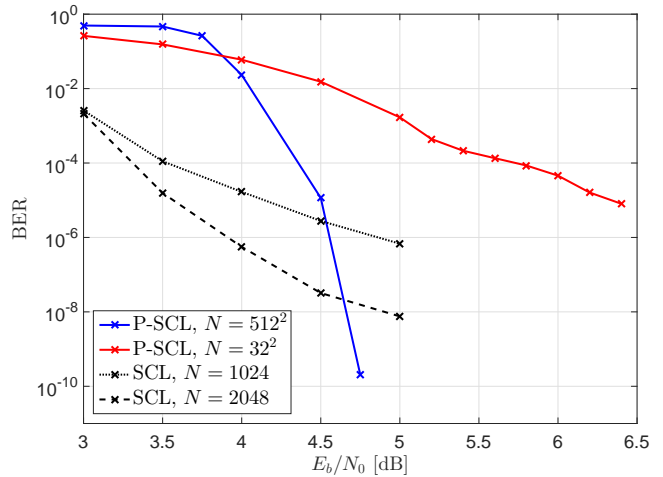}
  \caption{BER comparison for SCL and P-SCL, for codes of rate $R=(7/8)^2$ and $L=8$.}
  \label{fig:ECP-SCL}
\end{figure}

Figure~\ref{fig:ECP-SCL} depicts the BER curves for the same codes, obtained through SCL and P-SCL decoding with a list size $L=8$, and no CRC.
The more powerful SCL algorithm leads to an earlier waterfall region for all codes, with a slope slightly gentler than that of SC.
% Since P-SCL exploits SCL only for component codes, its improvement with respect to P-SC is lower than the gain obtained by SCL with respect to SC.
The P-SCL curve crosses the SCL ones around similar BER points as in Figure~\ref{fig:ECP-SC}, but at lower $E_b/N_0$.

%\begin{figure}
%  \centering
%   \scalebox{1}{\input{figures/ECP2.tikz}}
%   \ref{ECP-SCL}
%  \\
%  \vspace{2pt}
%  \caption{BER and BLER curves for SCL and P-SCL, for codes of rate $R=(7/8)^2$ and $L=8$.}
%  \label{fig:ECP-SCL}
%\end{figure}
%

%\fixme{Add comments on Figure~\ref{fig:ECP-OFC}, or remove it.}{}
%
%\begin{figure}
%  \centering
%   \scalebox{1}{\input{figures/ECP_OFC.tikz}}
%   \ref{ECP-OFC}
%  \\
%  \vspace{2pt}
%  \caption{BER curves for $N=65536$, $R=(239/256)^2$ for P-SC and P-SCL $L=16$, and the $(256^2,239^2)$ polar turbo product code proposed in \cite{KoikeAkinoIrregularPT}.}
%  \label{fig:ECP-OFC}
%\end{figure}

\subsection{Decoding Latency} \label{subsec:speed}

%SC-based decoding algorithms are inherently sequential: long codes thus suffer from very long decoding latency and low throughput. 
%While many techniques have been proposed to prune the SC decoding tree \cite{sarkis,hashemi_SSCL,hashemi_FSSCL}, \fixme{effective parallelization of SC-based decoding has not been achieved. }{Check if it's true}

\begin{figure}
  \centering
  \includegraphics[width=0.47\textwidth]{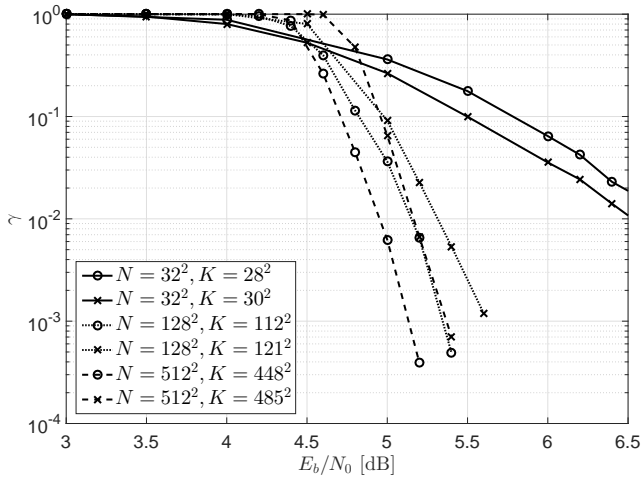}
  \caption{Evolution of $\gamma$ with codes of different length and rate, SC component decoding, $t=4$, $N_r=N_c$, $R_r=R_c$.}
  \label{fig:Gamma-SC}
\end{figure}

To begin with, we study the evolution of the parameters $\gamma$ and $t_{avg}$ in \eqref{eq:time} under SC decoding. 
Figure~\ref{fig:Gamma-SC} depicts the value of $\gamma$ measured at different $E_b/N_0$, for various code lengths and rates. 
%four code lengths and two code rates. 
The codes have been decoded with the proposed two-step decoding approach, considering $t = 4$ maximum iterations. 
%They assume an additive white Gaussian noise (AWGN) channel and binary phase-shift keying modulation. 
%\fixme{The frozen set $\mathcal{F}$ has been optimized for product decoding as described in Section \ref{sec:prodPol}. }{Add details, rephrase whatever based on Section \ref{subsec:F}}
As $E_b/N_0$ increases, the number of times SC is activated rapidly decreases towards $0$, with $\gamma<10^{-3}$ at a BER orders of magnitude higher than the working point for optical communications, which is the target scenario for the proposed construction. % (see Fig. \ref{fig:ECP-SC} in Section \ref{subsec:ECP}). 
Simulations have shown that the slope with which $\gamma$ tends to $0$ changes depending on the value of $t$; as $t$ increases, so does the steepness of the $\gamma$ curve. 
Regardless of $t$, $\gamma$ tends to $0$ as the channel conditions improve.

%\begin{figure}
%  \centering
%   \scalebox{1}{\input{figures/Gamma.tikz}}
%   \ref{Gamma-SC}
%  \\
%  \vspace{2pt}
%  \caption{Evolution of $\gamma$ with codes of different length and rate, SC component decoding, $t=4$, $N_r=N_c$, $R_r=R_c$. }
%  \label{fig:Gamma-SC}
%\end{figure}
%
%

%In (\ref{eq:timeSC}), it is assumed that the decoding process always performs $t$ iterations. 
The first decoding step is stopped as soon as $X_r=X_c$, or if the maximum number of iterations $t$ has been reached. 
Through simulation, we have observed that the average number of iterations $t_{avg}$ follows a behavior similar to that of $\gamma$, and tends to $1$ as $E_b/N_0$ increases.
It is worth noting that similar considerations apply when a decoding algorithm different than SC is used, as long as the same decoder is applied to the component codes and the length-$N$ code. 
The trends observed with SC for $\gamma$ and $t_{avg}$ are found with P-SCL as well, and we can safely assume that similar observations can be made with other SC-based decoding algorithms. 

%Assuming no restrictions of available resources, $\Delta^{SC}_N=2N-2$. 
%For example, for SCL decoding, $\Delta^{SCL}_N=2N+K-2$, and 
%\begin{align} \label{eq:timeSCL}
%\Delta^{\rm P-SCL}_N = &\left(\max(2N_r+K_r,2N_c+K_c)-2\right)\times t_{avg} +\nonumber\\
%&+ \gamma(2N+K-2)~.
%\end{align}
%In these cases, we can approximate the number of time steps required by product decoding of polar codes as
%\begin{equation}\label{eq:timeGeneral}
%\Delta^{\rm P}_N \approx \frac{\Delta_N}{\max(N_r,N_c)}~.
%\end{equation}
%When code-structure-based pruning algorithms are used \cite{sarkis,hashemi_SSCL,hashemi_FSSCL_TSP}, however, the relationship between $\Delta^{\rm P}_N$ and $\Delta_N$ is dependent on the frozen sets of the component codes and of the length-$N$ code, and can vary from (\ref{eq:timeGeneral}) significantly.

Table \ref{tab:speed} reports $\Delta_N$ required by standard SC and SCL decoders, as well as for the proposed two-step decoder P-SC and P-SCL, at different code lengths and rates. 
Assuming no restrictions of available resources, the number of time steps required by SC decoding is $\Delta^{\rm SC}_N=2N-2$, that becomes $\Delta^{\rm SCL}_N=2N+K-2$ for SCL decoding \cite{hashemi_FSSCL_TSP}.
For P-SC and P-SCL, $\Delta^{\rm P}_N$ is evaluated for worst case (WC), that assumes $t_{avg}=t$ and $\gamma=1$, and best case (BC), that assumes $t_{avg}=1$ and $\gamma=0$.
%For P-SC and P-SCL, worst case and best case $\Delta_N$ are given. 
%Worst case considers $t_{avg}=t$ and $\gamma=1$, while best case assumes $t_{avg}=1$ and $\gamma=1$. 
Simulation results show that $\Delta^{\rm P}_N$ tends to the asymptotic limit represented by BC decoding latency as the BER goes towards optical communication working point. 
As an example, for $N = 512^2 = 262144$, $K = 448^2 = 200704$ with P-SC, at $\text{BER} \simeq 2.5 \cdot 10^{-7}$, i.e. approximately eight orders of magnitude higher than the common target for optical communications, $\gamma \approx 6 \cdot 10^{-3}$ and $t_{avg}=1.1$, leading to $\Delta_N^{\rm P-SC}=5967$. 
This value is equivalent to $1.1\%$ of standard decoding time $\Delta_N^{\rm SC}$, while the BC latency is $0.2\%$ of $\Delta_N^{\rm SC}$. 
%This value is equivalent to $0.011\times\Delta_N^{SC}$, while the best case latency is $0.002\times\Delta_N^{SC}$. 
At $\text{BER} \simeq 10^{-15}$, it is safe to assume that the actual decoding latency is almost equal to BC.

\begin{table}
\centering
\scriptsize
%\caption{Time step analysis for standard and two-step decoding. Worst case (WC) considers $t_{avg}=4$, $\gamma=1$. Best case (BC) considers $t_{avg}=1$, $\gamma=0$.}
\caption{Time step analysis for standard and two-step decoding.}
\label{tab:speed}
\setlength{\extrarowheight}{1.5pt}
\begin{tabular}{c||c|cc||c|cc}

Code & \multirow{2}{*}{$\Delta^{\rm SC}_N$} & \multicolumn{2}{c||}{$\Delta^{\rm P-SC}_N$} & \multirow{2}{*}{$\Delta^{\rm SCL}_N$} & \multicolumn{2}{c}{$\Delta^{\rm P-SCL}_N$} \\ 

$N$,$K$ & & WC & BC & & WC & BC \\ 
\hline
\hline
$1024, 784$  &2046 & 2294 & 62 &2830 & 3190& 90 \\ 
$1024, 841$  &2046 & 2294 & 62 &2876 & 3240& 91 \\ 
$4096, 3136$  &8190 & 8694 & 126 &11326 & 12054& 182\\ 
$4096, 3249$ &8190 & 8694 &126  &11508 & 12244& 184\\ 
$16384, 12544$ &32766 & 33782 & 254&45310  & 46774& 366\\ 
$16384, 13225$ &32766 & 33782 &254 &46038 & 47518& 370\\ 
$65536, 50176$ &131070 & 133110 &510 &181246  & 184182& 734\\ 
$65536, 52900$ &131070 & 133110 &510 &184155  & 187119& 741\\ 
$262144, 200704$  &524286 & 528374 & 1022 &724990 & 730870& 1470\\ 
$262144, 211600$ &524286 & 528374 & 1022 &736623 & 742555& 1483\\ 
\end{tabular}
\end{table}

\section{Conclusion} \label{sec:conc}
In this paper, we have shown that the product of two non-systematic polar codes results in a polar code whose transformation matrix and frozen set are inferred from the component polar codes. 
%In this work, we have shown that product codes with non-systematic polar codes as component codes are polar codes. 
We have then proposed a code construction and decoding approach that exploit the dual nature of the resulting product polar code. 
%The frozen set of the polar codes is inferred from that of the component codes. 
The resulting code is decoded first as a product code, obtaining substantial latency reduction, while standard polar decoding is used as post-processing in case of failures. 
Performance analysis and simulations show that thanks to the high throughput of the proposed decoding approach, very long codes can be targeted, granting good error-correction performance suitable for optical communications.
%It is worth to notice that it is possible also to invert the construction order, namely taking a polar code and decompose it as the product of smaller polar codes. 
%This process is out of the scope of this paper and will be elaborated in a future paper. 
%We conjecture that the proposed product polar code construction can be inverted, namely rewriting any polar code as the product of smaller polar codes. 
Future works rely on the inversion of the proposed product polar code construction, namely rewriting any polar code as the product of smaller polar codes. 

\bibliographystyle{IEEEbib}
%\bibliography{IEEEabrv,refs}

\end{document}